\documentclass[aps,prd,onecolumn,groupedaddress,showpacs,nofootinbib,amssymb]{revtex4}
\usepackage{graphicx,bm,color}
\usepackage{amsmath}
\usepackage{amssymb}
\usepackage{amsfonts}

\newcommand{\be}{\begin{equation}}
\newcommand{\ee}{\end{equation}}
\newcommand{\bea}{\begin{eqnarray}}
\newcommand{\eea}{\end{eqnarray}}
\newcommand{\beaa}{\begin{eqnarray*}}
\newcommand{\eeaa}{\end{eqnarray*}}

\newcommand{\nn}{\nonumber \\}
\newcommand{\e}{\mathrm{e}}

\allowdisplaybreaks[4]

\begin{document}

\title{Renormalizable toy model of massive spin two field and 
new bigravity}

\author{Yuichi Ohara$^1$, Satoshi Akagi$^1$, and Shin'ichi Nojiri$^{1,2}$}

\affiliation{
$^1$ Department of Physics, Nagoya University, Nagoya
464-8602, Japan \\
$^2$ Kobayashi-Maskawa Institute for the Origin of Particles and
the Universe, Nagoya University, Nagoya 464-8602, Japan 
}

\begin{abstract}

In this paper, we propose a toy model of the renormalizable theory describing 
massive spin two field. 
Although the model is renormalizable, we show that the model contains ghost. 
The coupling of the theory with gravity can be regarded as a new kind of bimetric 
gravity or bigravity. We show that the massive spin two field plays the 
role of the cosmological constant. 

\end{abstract}


\maketitle

\section{Introduction \label{Sec1}}

After the establishment of the free theory of massive gravity by Fierz and Pauli 
\cite{Fierz:1939ix} (for a recent review, see \cite{Hinterbichler:2011tt}), any consistent 
interacting theory has not been found during three fourth centuries. 
One of the reasons is that there appears 
the Boulware-Deser ghost \cite{Boulware:1974sr,Boulware:1973my} 
in general and another is the appearance of the van Dam-Veltman-Zakharov (vDVZ) 
discontinuity \cite{vanDam:1970vg} in the massless limit, $m\to 0$ 
although the discontinuity can be screened 
by the Vainstein mechanism \cite{Vainshtein:1972sx} 
(see, for example, Ref.~\cite{Luty:2003vm}).

After the elapse of seventy five years from the work by Fierz and Pauli 
\cite{Fierz:1939ix}, there have been remarkable progress in the study of the non-linear 
massive gravity and the ghost-free models, which are  called the 
de Rham, Gabadadze, Tolley (dRGT) models, have been found  
in \cite{deRham:2010ik,deRham:2010kj,Hassan:2011hr}. 
The models have non-dynamical background metric but the models have been extended 
to the models with dynamical metric \cite{Hassan:2011zd,Hassan:2011vm,Hassan:2011tf}, 
which are called as bigravity models. 
%%%%%%%%%%%%%%%%%%%%%%%%%%%%%
After that, cosmology was studied in the massive gravity models \cite{deRham:2010tw} 
in the decoupling limit, where the models reduce to scalar-tensor theories 
and several activities in the massive gravity models 
\cite{Kluson:2012zz,Kluson:2012wf,Hassan:2011ea,D'Amico:2011jj} 
and in the bimetric gravity models 
\cite{Damour:2002wu,Volkov:2011an,vonStrauss:2011mq,
Berg:2012kn,Nojiri:2012zu,Nojiri:2012re,Bamba:2013fha,AKMS-TSK} 
have followed after that. 
%%%%%%%%%%%%%%%%%%%%%%%%%%%%%%%%%%%%%%%%%%

The absence of ghost was shown in the Hamiltonian analysis \cite{Hassan:2011tf} by using 
the Arnowitt-Deser-Misner (ADM) formalism, where the metric is assumed to be 
\be
\label{ADM}
g_{00} = - N^2+g^{ij}N_i N_{j}\, , \quad 
g_{0i} = N_{i}\, ,\quad g_{ij} = g_{ij}\, .
\ee
Here $i,j =1,2,3$ and $N$ is called as the lapse function and $N_i$ as the shift function. 
We denote the inverse of $g_{ij}$ by $g^{ij}$. 
In the dRGT models, after the redefinition of the shift function $N_i$, the Hamiltonian 
becomes linear to the lapse function $N$ and in the expression of the new shift function 
given by solving the equation obtained from the variation of the new shift function,  
the lapse function $N$ does not appear. 
Therefore the variation of $N$ give a constraint on $g_{ij}$ and their conjugate momenta. 
By combining this constraint with the secondary constraint derived from the constraint, 
an extra degree of freedom corresponding to the ghost is eliminated. 
%%%%%%%%%%%%%%%%%
Because the existence of the Boulware-Deser ghost may depend on initial conditions, 
the Boulware-Deser ghost in three dimensional bigravity model was studied in 
\cite{Banados:2013fda} by using the Hamiltonian analysis. 
%%%%%%%%%%%%%%%%

Recently in \cite{Hinterbichler:2013eza}, it has been proposed possibilities of new 
non-linear ghost-free derivative interactions in massive gravity. 
After that, however, in \cite{deRham:2013tfa}, it has shown that a class of the derivative 
interactions includes ghost and a kind of no-go theorem prohibiting the derivative 
interactions has been claimed. 
In this paper, we show the existence of the non-linear derivative interactions which are 
not included in \cite{deRham:2013tfa} although such derivative interactions generate 
ghost, unfortunately. 

%%%%%%%%%%%%%%
Motivated by such analyses, we propose a power counting renormalizable model 
describing the massive spin two field. The model could not be, however, really 
renormalizable because the projection operators included in the propagator 
generate non-renormalizble divergences. 
This problem is, however, solved by adding a term where a vector field couples 
with the massive spin two field. 
Although the model could be renormalizable, by investigating the spectrum of this model, 
we find that there could appear ghost and therefore the model cannot be realistic one but 
a kind of toy model. 
Because the gravity is not renormalizable, we may consider the coupling of the power 
counting renormalizable model, which could not be really renormalizable, with gravity. 
The model can be regarded as a new kind of bigravity. 
%%%%%%%%%%%%%

\section{Still New Derivative Interaction in Massive Gravity?}

In \cite{Hinterbichler:2013eza}, by using the perturbation $h_{\mu\nu}$ from the flat 
metric
\be
\label{h1}
h_{\mu\nu} = g_{\mu\nu} - \eta_{\mu\nu}\, ,
\ee
as a dynamical variable, new ghost free interactions were proposed. 
The interaction terms have the following form: 
\begin{equation}
\label{h2}
{\cal L}_{d, 0} \sim \eta^{\mu_{1} \nu_{1} \cdots \mu_{n} \nu_{n}} h_{\mu_{1} \nu_{1}} 
\cdots h_{\mu_{n} \nu_{n}}\, .
\end{equation}
or terms including $d$-derivative, which is called pseudo linear terms (see also 
\cite{Folkerts:2011ev}), 
\begin{equation}
\label{h3}
{\cal L}_{d, n} \sim \eta^{\mu_{1} \nu_{1} \cdots \mu_{n} \nu_{n}} \partial_{\mu_{1}} 
\partial_{\nu_{1}} h_{\mu_{2} \nu_{2}} \cdots \partial_{\mu_{d-1}} \partial _{\nu_{d-1}} 
h_{\mu_{d} \nu_{d}} h_{\mu_{d+1} \nu_{d+1}} \cdots h_{\mu_{n+d/2} nu_{n+d/2}} \, .
\end{equation}
Here $\eta^{\mu_{1} \nu_{1} \cdots \mu_{n} \nu_{n}}$ is given by the product of $n$ 
$\eta_{mu\nu}$ and anti-symmetrizing the indexes $\nu_1$, $\nu_2$, $\cdots$, and 
$\nu_n$, for examples
\begin{align}
\label{h3c}
\eta^{\mu_{1} \nu_{1} \mu_{2} \nu_{2}} \equiv &  
\eta^{\mu_{1} \nu_{1}} \eta^{\mu_{2} \nu_{2}} - \eta^{\mu_{1} \nu_{2}} 
\eta^{\mu_{2} \nu_{1}}\, , \nn
\eta^{\mu_{1} \nu_{1} \mu_{2} \nu_{2} \mu_{3} \nu_{3}} \equiv & 
\eta^{\mu_{1} \nu_{1}}\eta^{\mu_{2} \nu_{2}} \eta^{\mu_{3} \nu_{3}} 
 - \eta^{\mu_{1} \nu_{1}}\eta^{\mu_{2} \nu_{3}} \eta^{\mu_{3} \nu_{2}}
+ \eta^{\mu_{1} \nu_{2}}\eta^{\mu_{2} \nu_{3}} \eta^{\mu_{3} \nu_{1}} \nn
& - \eta^{\mu_{1} \nu_{2}}\eta^{\mu_{2} \nu_{1}} \eta^{\mu_{3} \nu_{3}}
+ \eta^{\mu_{1} \nu_{3}}\eta^{\mu_{2} \nu_{1}} \eta^{\mu_{3} \nu_{2}} 
 - \eta^{\mu_{1} \nu_{3}}\eta^{\mu_{2} \nu_{2}} \eta^{\mu_{3} \nu_{1}} 
\, .
\end{align}
It is evident that these terms are linear with respect to $h_{00}$, which could be a 
perturbation of the lapse function $N$ in the Hamiltonian and there do not appear the 
terms which include both of $h_{00}$ and $h_{0i}$. 
Therefore the variation of $h_{00}$ gives a constraint for $h_{ij}$ and their conjugate 
momenta $\pi_{ij}$ and 
%%%%%%%%
therefore the ghost could be eliminated although we may need more careful 
Hamiltonian analysis. 
%%%%%%%%%

The non-linear counterparts for (\ref{h2}) is nothing but the mass terms and the 
interaction terms in the dRGT models,
\be
\label{h3b}
\eta^{\mu_{1} \nu_{1} \cdots \mu_{n} \nu_{n}} h_{\mu_{1} \nu_{1}} \cdots h_{\mu_{n} \nu_{n}} 
\sim \sqrt{-g} g^{\mu_{1} \nu_{1} \cdots \mu_{n} \nu_{n}} {\cal K}_{\mu_{1} \nu_{1}} \cdots 
{\cal K}_{\mu_{n} \nu_{n} } \, .
\ee
Here $\mathcal{K}_{\mu\nu}$ is defined by 
\be
\label{h3d}
\mathcal{K}_\mu^{\ \nu} \equiv \delta_{\mu}^{\ \nu} - \sqrt{ g^{-1} f}_\mu^{\ \nu}\, ,
\ee
and $f_{\mu\nu}$ is the fiducial metric and often chosen to be $f_{\mu\nu} 
= \eta_{\mu\nu}$. 

In $D=4$ space-time dimensions, a possible non-trivial term with two derivative is given 
by
\begin{equation}
\label{h4}
{\cal L}_{2, 2}\sim \eta^{\mu_{1} \nu_{1} \mu_{2} \nu_{2} \mu_{3} \nu_{3}} 
\partial_{\mu_{1}} \partial_{\nu_{1}} h_{\mu_{2} \nu_{2}} h_{\mu_{2} \nu_{2}}\, ,
\end{equation}
and 
\begin{equation}
\label{h5}
{\cal L}_{2, 3}\sim \eta^{\mu_{1} \nu_{1} \mu_{2} \nu_{2} \mu_{3} \nu_{3}} 
\partial_{\mu_{1}} \partial_{\nu_{1}} h_{\mu_{2} \nu_{2}} h_{\mu_{2} \nu_{2}} h_{\mu_{3} \nu_{3}} 
\, .
\end{equation}
The non-linear counterpart of (\ref{h4}) could be nothing but the Einstein-Hilbert term. 
In case of the massive gravity, there is another candidate of the non-linear counterpart 
for (\ref{h4})  \cite{Kimura:2013ika}, which is 
\begin{equation}
\label{h6}
\sqrt{-g} g^{\mu_{1} \nu_{1} \mu_{2} \nu_{2} \mu_{3} \nu_{3}}
R_{\mu_1 \mu_{2} \nu_1 \nu_2} {\cal K}_{\mu_3 \nu_{3}} \, .
\end{equation}
The non-trivial, fully non-linear counterpart of (\ref{h5}) could be also given by 
\begin{equation}
\label{h7}
\sqrt{-g} g^{\mu_{1} \nu_{1} \mu_{2} \nu_{2} \mu_{3} \nu_{3} \mu_{4} \nu_{4}} 
R_{\mu_{1} \mu_{2} \nu_{1} \nu_{2}} {\cal K}_{\mu_{3} \nu_{3}} {\cal K}_{\mu_{4} \nu_{4} }\, .
\end{equation}
Here $g^{\mu_{1} \nu_{1} \cdots \mu_{n} \nu_{n}}$ is, as in the definition of 
$\eta^{\mu_{1} \nu_{1} \cdots \mu_{n} \nu_{n}}$, given by the product of $n$ $g_{\mu\nu}$ 
and anti-symmetrizing the indexes $\nu_1$, $\nu_2$, $\cdots$, and $\nu_n$. 

In \cite{deRham:2013tfa}, however, it has been shown that the non-linear terms 
(\ref{h6}) and (\ref{h7}) could generate the ghost by using the mini-superspace where
\be
\label{h8}
N=N(t)\, ,\quad N_i=0\, ,\quad g_{ij} = a(t)^2 \eta_{ij}\, .
\ee
In fact, in the mini-superspace (\ref{h8}), the terms (\ref{h6}) and (\ref{h7}) have the 
following form
\cite{deRham:2013tfa}:
\begin{align}
\label{h12}
\sqrt{-g} g^{\mu_{1} \nu_{1} \mu_{2} \nu_{2} \mu_{3} \nu_{3}}
R_{\mu_1 \mu_{2} \nu_1 \nu_2} {\cal K}_{\mu_3 \nu_{3}} \sim &
Na^3 \left[2\frac{\dot{a}^2}{a^2N^2}-\frac{\dot{a}^2}{a^3 N^2}+\frac{\dot{a}^2}{a^2 N^3} 
\right]\, , \\
\label{h13}
\sqrt{-g} g^{\mu_{1} \nu_{1} \mu_{2} \nu_{2} \mu_{3} \nu_{3} \mu_{4} \nu_{4}} 
R_{\mu_{1} \mu_{2} \nu_{1} \nu_{2}} {\cal K}_{\mu_{3} \nu_{3}} {\cal K}_{\mu_{4} \nu_{4} } \sim 
& N a^3 \left[ \frac{{\dot a}^2}{a^2 N^3} - \frac{{\dot a}^2}{a^3 N^2} 
+ \frac{{\dot a}^2}{a^2 N^3} -  \frac{{\dot a}^2}{a^3 N^3}\right]\, .
\end{align}
The expressions (\ref{h12}) and (\ref{h13}) tell that in the Hamiltonian, the terms 
(\ref{h6}) and (\ref{h7}) generate the terms which are not linear with respect to the lapse 
function $N$. 
Therefore the equation given by the variation of $N$ can be solved with respect to $N$ 
and does not give any constraint on $g_{ij}$ or their conjugate momenta, which tells that 
the ghost could not be eliminated. 

We should note that the terms (\ref{h6}) and (\ref{h7}) are not unique terms 
reproducing (\ref{h4}) and (\ref{h5}), respectively. 
Another candidate reproducing (\ref{h4}) is 
\be
\label{h14}
\sqrt{-g} g^{\mu_{1} \nu_{1} \mu_{2} \nu_{2} \mu_{3} \nu_{3}}
\left( \nabla_{\mu_1} \nabla_{\nu_1} \mathcal{K}_{\mu_2 \nu_2} \right) 
\mathcal{K}_{\mu_3 \nu_3}\, ,
\ee
and a candidate for (\ref{h5}) is
\be
\label{h15}
\sqrt{-g} g^{\mu_{1} \nu_{1} \mu_{2} \nu_{2} \mu_{3} \nu_{3} \mu_{4} \nu_{4}} 
\left( \nabla_{\mu_1} \nabla_{\nu_1} \mathcal{K}_{\mu_2 \nu_2} \right) 
\mathcal{K}_{\mu_3 \nu_3} \mathcal{K}_{\mu_4 \nu_4}\, .
\ee
In the mini-superspace (\ref{h8}), these terms can be expressed as
\begin{align}
\label{h16}
\sqrt{-g} g^{\mu_{1} \nu_{1} \mu_{2} \nu_{2} \mu_{3} \nu_{3}}
\left( \nabla_{\mu_1} \nabla_{\nu_1} \mathcal{K}_{\mu_2 \nu_2} \right) 
\mathcal{K}_{\mu_3 \nu_3} \sim & Na^3 \left[ \frac{{\dot a}^2}{a^2 N^4} \right]\, , \\
\label{h17}
\sqrt{-g} g^{\mu_{1} \nu_{1} \mu_{2} \nu_{2} \mu_{3} \nu_{3} \mu_{4} \nu_{4}} 
\left( \nabla_{\mu_1} \nabla_{\nu_1} \mathcal{K}_{\mu_2 \nu_2} \right) 
\mathcal{K}_{\mu_3 \nu_3} \mathcal{K}_{\mu_4 \nu_4} 
\sim & N a^3 \left[ \frac{{\dot a}^2}{a^3 N^4} - \frac{{\dot a}^2}{a^2 N^4}\right]\, .
\end{align}
 From the above expressions (\ref{h16}) and (\ref{h17}), however, we find that these 
terms (\ref{h14}) and (\ref{h15}) could also generate the ghost. 
The ghost could not be eliminated even if we consider the combinations in (\ref{h12}), 
(\ref{h13}), (\ref{h14}), (\ref{h16}), and (\ref{h17}). 

We should note that there is another candidate to reproduce (\ref{h4}):
\be
\label{h18}
\sqrt{-g} g^{\nu \nu^{\prime} \rho \rho^{\prime} \sigma \sigma^{\prime}} 
f^{\nu^{\prime} \nu^{\prime \prime}} \nabla_{\nu} \mathcal{K}_{\rho \rho^{\prime}} 
\nabla_{\nu^{\prime \prime} }\mathcal{K}_{\sigma \sigma^{\prime}} \, .
\ee
Here $f^{\mu\nu}=\eta^{\mu\nu}$. In the mini-superspace, however, this term has the 
following form:
\be
\label{h19}
\sqrt{-g} g^{\nu \nu^{\prime} \rho \rho^{\prime} \sigma \sigma^{\prime}} 
f^{\nu^{\prime} \nu^{\prime \prime}} \nabla_{\nu} \mathcal{K}_{\rho \rho^{\prime}} 
\nabla_{\nu^{\prime \prime} }\mathcal{K}_{\sigma \sigma^{\prime}} \sim Na^3 
\left[-\frac{6\dot{a}^2}{a^3N}+\frac{6\dot{a}^2}{aN^3}-\frac{6\dot{a}^2}{N^4}\right]\, ,
\ee
and therefore the ghost could not be eliminated even if we consider any combination 
with other terms. 

\newcommand{\invI}{{\bf I}^{-1}}   
\newcommand{\K}{\mathcal{K}}
\newcommand{\I}{\mathbf{I}}

Then we consider the possibility of other classes of the no-ghost interactions by relaxing 
the assumption in \cite{deRham:2013tfa}.   
In the argument so far, we have considered the terms which have invariance under the 
general coordinate transformation if the fiducial metric $f_{\mu\nu}$ could be a dynamical 
tensor. 
We may relax this condition and require only the Lorentz invariance. 
Then we may consider the term given by replacing the covariant derivatives 
$\nabla_\mu$ in (\ref{h18}) by the partial derivative $\partial_\mu$: 
\be
\label{last}
\sqrt{-g} g^{\nu \nu^{\prime} \rho \rho^{\prime} \sigma \sigma^{\prime}} 
f^{\nu^{\prime} \nu^{\prime \prime}} \partial_{\nu} 
\mathcal{K}_{\rho \rho^{\prime}} \partial_{\nu^{\prime \prime} }
\mathcal{K}_{\sigma \sigma^{\prime}} \, .
\ee
In the mini-superspace (\ref{h8}), this term is surely linear with respect to $N$. 
Then we now check if the term (\ref{last}) could give interactions without ghost 
by using the full ADM formalism. 
Explicitly the term (\ref{last}) has the following form:
\begin{align}
&\sqrt{-g}{\delta_{\mu_1}}^{[\nu_1} {\delta_{\mu_2}}^{\nu_2} {\delta_{\mu_3}}^{\nu_3]} 
\eta^{\mu_1 \rho} \partial_{\nu_1} {\K^{\mu_2}}_{\nu_2} \partial_{\rho} {\K^{\mu_3}}_{\nu_3}  
\nn 
&=\sqrt{-g} \left[ -\left(\partial_0 {\sqrt{g^{-1} \eta}^i}_i \right)^2 
+2 \left(\partial_i {\sqrt{g^{-1} \eta}^0}_0 \right) \left(\partial_i {\sqrt{g^{-1} \eta}^k}_k  
\right) +\left(\partial_i {\sqrt{g^{-1} \eta}^j}_j \right) 
\left(\partial_i {\sqrt{g^{-1} \eta}^k}_k \right)  \right. \nn
&-\left(\partial_0 {\sqrt{g^{-1} \eta}^0}_i \right) \left(\partial_i {\sqrt{g^{-1} \eta}^j}_j 
\right) +\left(\partial_i {\sqrt{g^{-1} \eta}^i}_0 \right) 
\left(\partial_0 {\sqrt{g^{-1} \eta}^j}_j \right)  -\left(\partial_i {\sqrt{g^{-1} \eta}^i}_j \right) 
\left(\partial_j {\sqrt{g^{-1} \eta}^0}_0 \right)  \nn
&-\left(\partial_i {\sqrt{g^{-1} \eta}^i}_j \right) \left(\partial_j {\sqrt{g^{-1} \eta}^k}_k 
\right) +\left(\partial_0 {\sqrt{g^{-1} \eta}^0}_i \right) 
\left(\partial_j {\sqrt{g^{-1} \eta}^i}_j \right) +\left(\partial_i {\sqrt{g^{-1} \eta}^i}_0 \right) 
\left(\partial_j {\sqrt{g^{-1} \eta}^0}_j \right) \nn
&-\left(\partial_i {\sqrt{g^{-1} \eta}^i}_j \right) \left(\partial_0 {\sqrt{g^{-1} \eta}^j}_0 
\right)+\left(\partial_i {\sqrt{g^{-1} \eta}^i}_j \right) 
\left(\partial_k {\sqrt{g^{-1} \eta}^j}_k \right) 
+\left(\partial_0 {\sqrt{g^{-1} \eta}^i}_j \right) \left(\partial_0 {\sqrt{g^{-1} \eta}^j}_i \right) 
\nn
&-2\left(\partial_i {\sqrt{g^{-1} \eta}^0}_j \right) \left(\partial_i {\sqrt{g^{-1} \eta}^j}_0 
\right)  -\left(\partial_i {\sqrt{g^{-1} \eta}^j}_k \right) 
\left(\partial_i {\sqrt{g^{-1} \eta}^k}_j \right)+\left(\partial_0 {\sqrt{g^{-1} \eta}^i}_j \right) 
\left(\partial_j {\sqrt{g^{-1} \eta}^0}_i \right) \nn
&+\left(\partial_i {\sqrt{g^{-1} \eta}^0}_j \right) \left(\partial_j {\sqrt{g^{-1} \eta}^i}_0 
\right) -\left(\partial_i {\sqrt{g^{-1} \eta}^j}_0 \right) 
\left(\partial_0 {\sqrt{g^{-1} \eta}^i}_j \right) 
+\left(\partial_i {\sqrt{g^{-1} \eta}^j}_k \right) \left(\partial_k {\sqrt{g^{-1} \eta}^i}_j \right) 
\nn
&-\left(\partial_0 {\sqrt{g^{-1} \eta}^j}_j \right) \left(\partial_i {\sqrt{g^{-1} \eta}^0}_i 
\right) +\left(\partial_i {\sqrt{g^{-1} \eta}^j}_j \right) 
\left(\partial_0 {\sqrt{g^{-1} \eta}^i}_0 \right) -\left(\partial_i {\sqrt{g^{-1} \eta}^0}_0 
\right) \left(\partial_j {\sqrt{g^{-1} \eta}^i}_j \right) \nn
&\left.-\left(\partial_i {\sqrt{g^{-1} \eta}^k}_k \right) \left(\partial_j {\sqrt{g^{-1} \eta}^i}_j 
\right) \right] \, .
\label{lastadm}
\end{align}
In order that ghost could not appear, the term should be given in the form where the 
time-derivative of the lapse and shift functions do not appear. 
This kind of form might be obtained by the cancellations between several terms after the 
partial integration. 
Because this kind of the cancellation should occur between the terms including the same 
number of the time derivatives, we now consider the following terms:
\begin{align}
\sqrt{-g} \left[-\left(\partial_0 {\sqrt{g^{-1} \eta}^i}_i \right)^2  
+\left(\partial_0 {\sqrt{g^{-1} \eta}^i}_j \right) \left(\partial_0 {\sqrt{g^{-1} \eta}^j}_i \right)    
\right] 
\label{lastadm2}
\end{align}
As in \cite{Hassan:2011hr}, for convenience, we use the redefined shift function $n^i$, 
which is given by 
\be
N^i = (\delta^i_j + N D^i_{\,\,j}) n^j \, .
\label{newshift}
\ee
The definition of $D^i_{\,\,j}$ is given by solving the following equation \cite{Hassan:2011hr} 
\be
\label{h20}
(\sqrt{1-n^T\I\, n})\, D=\sqrt{(\gamma^{-1}-Dn n^TD^T)\I} \, , \quad 
\I=\delta_{ij}\, , \quad \invI=\delta^{ij}\, .
\ee
By using $n^i$, we rewrite $\sqrt{g^{-1}\eta}^{\mu}_{\ \nu}$ as follows,
\be
\label{h21}
\sqrt{g^{-1} \eta} = \frac{1}{N}\mathcal{A}+\mathcal{B}\, .
\ee
Here
\begin{align}
\label{h22}
&\mathcal{A}=\frac{1}{\sqrt{1-n^{T} \bm{I} n}}\left(
\begin{array}{cc}
1 & n^{T} \bm{I} \\
-n & -n n^{T} \bm{I}
\end{array}
\right) \quad 
\mathcal{B}=\left(
\begin{array}{cc}
0 & 0 \\
0 & \sqrt{(\gamma^{-1}-Dnn^{T}D^{T})\bm{I}}
\end{array} 
\right) \, .
\end{align}
In order to simplify the notation, we define the following quantities:
\be
A:=\frac{1}{\sqrt{1-n^{T} \bm{I} n}}\, ,\quad B^{l} :=n^{l}\, , \quad 
{C^i}_j:=\sqrt{(\gamma^{-1}-Dnn^{T}D^{T})\bm{I}} \, .
\label{newvar}
\ee
By using (\ref{newvar}), 
$\sqrt{g^{-1}\eta}^{\mu}_{\ \nu}$ can be rewritten as
\be
\label{h24}
\sqrt{g^{-1} \eta}^{\mu}_{\ \nu}=\left(
\begin{array}{cc}
A/N & AB^{l}\delta_{li}/N \\
 -AB^j / N & -B^i B^k \delta_{kj}/N +{C^{i}}_{\ j} 
\end{array}
\right)\, ,
\ee
and $\partial_0 \sqrt{g^{-1}\eta}^{i}_{\ j}$ can be expressed as follows,
\be
\partial_0 {\sqrt{g^{-1}\eta}^{i}}_{j} = \frac{B^i B^k \delta_{kj} \dot{N}}{N^2} 
-\frac{\dot{B}^i B^k \delta_{kj}}{N}-\frac{B^i \dot{B}^k \delta_{kj}}{N}+{\dot{C}^{i}}_{\ j} \, .
\label{psqrtm}
\ee
Therefore by using ADM variables, Eq.~ (\ref{lastadm2}) has the following form: 
\begin{align}
&\sqrt{-g} \left[-\left(\partial_0 {\sqrt{g^{-1} \eta}^i}_i \right)^2  
+\left(\partial_0 {\sqrt{g^{-1} \eta}^i}_j \right) \left(\partial_0 {\sqrt{g^{-1} \eta}^j}_i \right)    
\right]  \nn
&=N\sqrt{\gamma}\left[-\frac{2(B^i \dot{B}^k \delta_{ik})^2}{N^2} -({\dot{C}^{i}}_{\ i})^2 
+\frac{4(B^i \dot{B}^k \delta_{ik}) {\dot{C}^{j}}_{\ j}}{N} - \frac{2(B^i B^k \delta_{ik})\dot{N} 
{\dot{C}^{j}}_{\ j}}{N^2} \right. \nn
&\left.+\frac{2B^i B^k \delta_{kj} {\dot{C}^{j}}_{\ i} \dot{N}}{N^2} 
+ \frac{2(B^l \delta_{li} B^i)(\dot{B}^l \delta_{li} \dot{B}^i)}{N^2}-\frac{2\dot{B}^i B^k 
\delta_{kj} {\dot{C}^{j}}_{\ i}}{N} 
 -\frac{2 B^i \dot{B}^k \delta_{kj} {\dot{C}^{j}}_{\ i}}{N}+\dot{C}^{i}_{\ j} \dot{C}^{j}_{\ i} 
\right] \, .
\label{lastadm2b}
\end{align}
From the expression (\ref{lastadm2b}), we find the time-derivatives of the lapse and shift 
functions cannot be canceled and therefore there could appear ghost. 

\section{Renormalizable model of massive spin two field}

We now propose a power-counting renormalizable 
model of the massive spin two field, whose Lagrangian density is given by
\begin{align}
\label{hh10}
\mathcal{L}_{h0} 
= & - \frac{1}{2} \eta^{\mu_{1} \nu_{1} \mu_{2} \nu_{2} \mu_{3} \nu_{3}} 
\left( \partial_{\mu_{1}} \partial_{\nu_{1}} h_{\mu_{2} \nu_{2}}\right) h_{\mu_{3} \nu_{3}}
+ \frac{m^2}{2} \eta^{\mu_{1} \nu_{1} \mu_{2} \nu_{2}} h_{\mu_{1} \nu_{1}} 
h_{\mu_{2} \nu_{2}} \nn
& - \frac{\mu}{3!} \eta^{\mu_{1} \nu_{1} \mu_{2} \nu_{2} \mu_{3} \nu_{3}} 
h_{\mu_{1}\nu_{1}} h_{\mu_{2} \nu_{2}} h_{\mu_{3} \nu_{3}}
 - \frac{\lambda}{4!} \eta^{\mu_{1} \nu_{1} \mu_{2} \nu_{2} \mu_{3} \nu_{3} \mu_{4} \nu_{4}} 
h_{\mu_{1} \nu_{1}} h_{\mu_{2} \nu_{2}} h_{\mu_{3} \nu_{3}} h_{\mu_{4} \nu_{4}}
\, \nn
= & - \frac{1}{2} \left( h \Box h - h^{\mu\nu} \Box h_{\mu\nu} 
 - h \partial^\mu \partial^\nu h_{\mu\nu} - h_{\mu\nu} \partial^\mu \partial^\nu h 
+ 2 h_\nu^{\ \rho} \partial^\mu \partial^\nu h_{\mu\rho} \right) \nn
& + \frac{m^2}{2} \left( h^2 - h_{\mu\nu} h^{\mu\nu} \right) 
 - \frac{\mu}{3!} \left( h^3 - 3 h h_{\mu\nu} h^{\mu\nu} 
+ 2 h_\mu^{\ \nu} h_\nu^{\ \rho} h_\rho^{\ \mu} \right)  \nn
& - \frac{\lambda}{4!} \left( h^4 - 6 h^2 h_{\mu\nu} h^{\mu\nu} 
+ 8 h h_\mu^{\ \nu} h_\nu^{\ \rho} h_\rho^{\ \mu} 
 - 6 h_\mu^{\ \nu} h_\nu^{\ \rho} h_\rho^{\ \sigma} h_\sigma^{\ \mu} 
+ 3 \left( h_{\mu\nu} h^{\mu\nu} \right)^2\right) \, .
\end{align}
Here $m$ and $\mu$ are parameters with the dimension of mass and $\lambda$ is 
a dimensionless parameters. Therefore the model given by the Lagrangian is 
power-counting renormalizable. 
%%%%%%%%%%%%%%%%%%%%
The model could be also free from ghost. 

We should note, however, that the propagator is given by 
\begin{align}
\label{hh1}
D^m_{\alpha\beta,\rho\sigma} =& - \frac{1}{2\left( p^2 + m^2 \right)}
\left\{  P^m_{\alpha\rho} P^m_{\beta\sigma} + P^m_{\alpha\sigma} P^m_{\beta\rho}  
 - \frac{2}{3} P^m_{\alpha\beta} P^m_{\rho\sigma} \right\} \, , \\
\label{hh2}
P^m_{\mu\nu} \equiv& \eta_{\mu\nu} + \frac{p_\mu p_\nu}{m^2} \, .
\end{align}
Then when $p^2$ is large, the propagator behaves as  
$D^m_{\alpha\beta,\rho\sigma} \sim \mathcal{O}\left( p^2 \right)$ 
due to the projection operator $P^m_{\mu\nu}$, which makes the behavior for large
 $p^2$ worse and therefore the model should not be renormalizable. 

There is a similar problem in the model of massive vector field, whose Lagrangian density 
is given by
\be
\label{hh3}
\mathcal{L} = -\frac{1}{4} \left(\partial_\mu A_\nu - \partial_\nu A_\mu \right)
\left(\partial^\mu A^\nu - \partial^\nu A^\mu \right) 
 - \frac{1}{2}m^2 A^\mu A_\mu \, .
\ee
The propagator $D_{\mu\nu}$ of the massive vector is given by
\be
\label{hh4}
D_{\mu\nu} = - \frac{1}{ p^2 + m^2 }P^m_{\mu\nu}\, ,
\ee
which is the inverse of 
\be
\label{hh5}
O^{\mu\nu} \equiv - \left( p^2 + m^2 \right) \eta^{\mu\nu} + p^\mu p^\nu\, ,
\ee
that is
\be
\label{hh6}
O^{\mu\nu}D_{\nu\rho} = \delta^\mu_{\ \rho}\, .
\ee
The expression (\ref{hh4}) tells that for  large $p^2$, $D_{\mu\nu}$ behaves 
as $\mathcal{O}(1)$ and therefore the model (\ref{hh3}) could not be renormalizable. 
If the vector field, however, couples only with the conserved current $J_\mu$ 
which satisfies the conservation law $\partial^\mu J_\mu =0$, the term 
$\frac{p_\mu p_\nu}{m^2}$ in the projection operator $P^m_{\mu\nu}$ drops and 
the propagator behaves as $D_{\mu\nu} \sim \mathcal{O}\left(1/p^2\right)$ 
and therefore the model may become renormalizable. 

Instead of imposing the conservation law, we may add the following term to the action:
\be
\label{hh7}
2 \alpha \phi \partial^\mu A_\mu \, ,
\ee
and consider the inverse of the operator 
\be
\label{hh8}
O_{A\phi} = \left( \begin{array}{cc}
O^{\mu\nu} & -i \alpha p^\mu \\
i \alpha p^\nu & 0 \end{array}
\right)\, ,
\ee
which is given by
\begin{align}
\label{hh9}
D_{A\phi} =& \left( \begin{array}{cc}
 - \frac{1}{p^2 + m^2} P_{\nu\rho} & - i \frac{p_\nu}{\alpha p^2} \\
i \frac{p_\rho}{\alpha p^2} & \frac{m^2}{\alpha^2 p^2} \end{array} \right)\, , \\
\label{R3}
P^{\mu\nu} \equiv& \eta^{\mu\nu} - \frac{p^\mu p^\nu}{p^2} \, , \\
\label{hh10b}
O_{A\phi}  D_{A\phi} =& \left( \begin{array}{cc}
\delta^\mu_{\ \rho} & 0 \\
0 & 1 \end{array} \right)\, .
\end{align}
The projection operator $P_{\mu\nu}$ is equal to the projection operator 
$P^m_{\mu\nu}$ on shell, $p^2 = -m^2$, but the behavior for large $p^2$ becomes 
different from each other. 
As a result, the propagator between two $A_\mu$'s behaves as 
$\mathcal{O}\left(1/p^2\right)$ and therefore the model could become 
renormalizable if the 
interaction terms are also renormalizable. 
%%%%%%%%%%%%
We should note that by construction, we are assuming that the interactions are given by 
$A_\mu$ and the interactions do not include the scalar field $\phi$. This tells that in the 
internal lines of the loops in the Feynmann diagrams, the propagators of the two vector 
fields $A$ appear but the propagators between two scalars $\phi$ nor those between the 
vector field $A_\mu$ and the scalar field $\phi$ do not appear. 
Therefore although the propagator between the vector field $A_\mu$ and the 
scalar field $\phi$ behaves as $\mathcal{O}\left(1/p\right)$ instead of 
$\mathcal{O}\left(1/p^2\right)$, this behavior could not generate non-renormalizable 
divergence. 

As we will see, however, the term (\ref{hh7}) generates a ghost. 
The total Lagrangian density (\ref{hh3}) with (\ref{hh7}) can be diagonalized by redefining 
the vector field $A_\mu$ by a new vector field $B_\mu$, which is given by 
\be
\label{RR1}
A_\mu = B_\mu - \frac{2\alpha}{m^2}\partial_\mu \phi\, ,
\ee
and we obtain
\be
\label{RR2}
\mathcal{L} = -\frac{1}{4} \left(\partial_\mu B_\nu - \partial_\nu B_\mu \right)
\left(\partial^\mu B^\nu - \partial^\nu B^\mu \right) 
 - \frac{1}{2}m^2 B^\mu B_\mu 
+ \frac{2\alpha^2}{m^2}\partial^\mu \phi \partial_\mu \phi\, . 
\ee
The propagator of the redefined vector field is given by (\ref{hh4}) and therefore this 
propagator might appear to generate non-renormalizable divergences. 
We should also note that there appear non-renormalizable derivative interactions of the 
scalar field, which include $\partial_\mu \phi$. 
The non-renormalizable divergences generated by the derivative interactions should be 
canceled by the non-renormalizable divergences coming from the propagator 
(\ref{hh4}) of the redefined vector field $B_\mu$ and there could remain only 
renormalizable divergences. 
The cancellation is consistent with the renormalizability given by the propagator in 
(\ref{hh9}).  
An important point is the following: We assume, by construction, that the interactions are 
not given in terms of the redefined vector field $B_\mu$ but in terms of $A_\mu$, which 
is the vector field before the redefinition (\ref{RR1}) and the interactions do not include 
the scalar field $\phi$, either. 
Therefore in the internal lines of the loops in the Feynmann diagrams, the propagators of 
the two vector fields always appear in the form of the propagators between the two 
vector fields $A_\mu$ in (\ref{hh9}) and therefore there could not appear 
non-renormalizable divergences coming from the projection 
operator (\ref{hh2}) in the propagator (\ref{hh4}). 

We should note, however, the $+$ sign in front of the kinetic term of the 
scalar field tells that the scalar field is ghost, which generates the negative norm 
states in the quantum theory and therefore the model given here is not consistent as a 
quantum theory.  
%%%%%%%%%%%%%

Anyway we may consider deformation similar to (\ref{hh7}) of the model by adding the 
following new term to the Lagrangian density (\ref{hh10}):
\be
\label{hhh10}
\mathcal{L} = \mathcal{L}_{h0} + 4\alpha A^\mu \partial^\nu h_{\mu\nu}\, ,
\ee
and consider the following equation: 
\be
\label{R1}
\left( \begin{array}{cc}
\mathcal{O}^{\mu\nu,\alpha\beta} & 
 -i \alpha \left( p^\mu \eta^{\alpha\nu} + p^\nu \eta^{\alpha\mu} \right) \\
i \alpha \left( p^\alpha \eta^{\mu\beta} + p^\beta \eta^{\mu\alpha} \right) & 0 
\end{array} \right)
\left( \begin{array}{cc}
D_{\alpha\beta,\rho\sigma} & - i E_{\sigma,\alpha\beta} \\
 i E_{\alpha,\rho\sigma} & F_{\alpha\sigma} 
\end{array} \right) 
= \left( \begin{array}{cc} \frac{1}{2} \left( \delta^\mu_{\ \rho} \delta^\nu_{\ \sigma} 
+ \delta^\mu_{\ \alpha} \delta^\nu_{\ \beta} \right) & 0 \\
0 & \delta^\mu_{\ \sigma} \end{array} \right) \, .
\ee
We should note that $\mathcal{O}^{\mu\nu,\alpha\beta}$ 
is given by
\begin{align}
\label{R2}
\mathcal{O}^{\mu\nu,\alpha\beta} = & 
 - \left\{ \frac{1}{2} \left( P^{\mu\alpha} P^{\nu\beta} 
+ P^{\mu\beta} P^{\nu\alpha} \right) - P^{\mu\nu} P^{\alpha\beta} \right\} 
\left( p^2 + m^2 \right) \nn
& - \left\{ \frac{1}{2} \left( p^\alpha p^\mu P^{\nu\beta} 
+ p^\beta p^\mu P^{\nu\alpha} + p^\alpha p^\nu P^{\mu\beta} 
+ p^\beta p^\nu P^{\mu\alpha} \right) - p^\mu p^\nu P^{\alpha\beta} 
 - p^\alpha p^\beta P^{\mu\nu}
\right\} \frac{m^2}{p^2} \, .
\end{align}
Then we find 
\begin{align}
\label{R4}
D_{\alpha\beta,\rho\sigma} 
= & -\frac{1}{2\left( p^2 + m^2 \right)} \left\{ P_{\alpha\rho} P_{\beta\sigma} 
+ P_{\alpha\sigma} P_{\beta\rho} - P_{\alpha\beta} P_{\rho\sigma} \right\} \, , \\
\label{R5}
E_{\alpha,\rho\sigma} = & \frac{1}{2\alpha p^2} \left\{ p_\rho P_{\alpha\sigma} 
+ p_\sigma P_{\alpha\rho} - \frac{m^2 p_\alpha}{2 \left( p^2 + m^2 \right)} P_{\rho\sigma} 
+ \frac{p_\alpha p_\rho p_\sigma}{p^2} \right\}\, , \\
\label{R6}
F_{\alpha\sigma} = & \frac{m^2}{2 \alpha^2 p^2} P_{\alpha\sigma} 
+ \frac{3 m^4}{8\alpha^2 \left( p^2 \right)^2 
\left( p^2 + m^2 \right)} p_\alpha p_\sigma \, .
\end{align}
Because the propagator between two $h_{\mu\nu}$'s behaves 
as $\mathcal{O}\left(1/p^2\right)$, the model could become renormalizable. 

We should note that the coupling of $h_{\mu\nu}$ with the energy-momentum tensor 
$T_{\mu\nu}$, $\kappa^2 h^{\mu\nu} T_{\mu\nu}$, which appears in the general relativity, 
breaks the renormalizability because $\kappa$ has the dimension of length. 
The coupling with a scalar field $\phi$ or the Rarita-Schwinger field $\psi_\mu$ can be, 
however, renormalizable, 
\be
\label{h11}
\phi \eta^{\mu_{1} \nu_{1} \mu_{2} \nu_{2} \mu_{3} \nu_{3}} \, , \quad 
h^{\mu\nu} {\bar \psi}_\mu \psi_\nu\, ,
\ee
which may appear when we supersymmetrize the action (\ref{hh10}) or (\ref{hhh10}). 

\section{Hamiltonian analysis and spectrum}

It is not so clear what could be physical degrees of freedom in the Lagrangian 
(\ref{hhh10}). 
Then in this section, we count the physical degrees of freedom by using the Hamiltonian 
analysis (for example, see \cite{Hinterbichler:2011tt}). 
After that, we diagonalize the free part and find what could  be the physical degrees of 
freedom. 

The free part $\mathcal{L}_0$ of the Lagrangian (\ref{hhh10}) is given by
\begin{align}
\label{OAN1}
\mathcal{L}_0 = -\frac{1}{2} \partial_\lambda h_{\mu \nu} \partial^{\lambda} h^{\mu \nu} 
+\partial_\mu h_{ \nu \lambda } \partial^{\nu} h^{\mu \lambda } 
 - \partial_\mu h^{\mu \nu} \partial_\nu h 
+ \frac12 \partial_\lambda h \partial^\lambda h -\frac12 m^2 (h_{\mu \nu} h^{\mu \nu} 
 - h^2) + 4 \alpha A_\mu \partial_\nu h^{\mu \nu} \, .
\end{align}
We now investigate the structure of the constraints for the free Lagrangian (\ref{OAN1}). 
By the integration by part, the free Lagrangian (\ref{OAN1}) can be rewritten as follows, 
\begin{align}
\label{OAN2}
\mathcal{L}_0 =& \mathcal{F}(h_{ij} , \dot{h}_{ij} , h_{0i}) + h_{00} \mathcal{G} (h_{ij}) 
+ 2\alpha A_\mu \partial_\nu h^{\mu \nu } 
 -2\alpha \partial_{(\mu} A_{\nu )}h^{\mu \nu}  \, ,\\ 
\label{OAN2B}
\mathcal{F}(h_{ij} , \dot{h}_{ij} , h_{0i}) 
=& \frac12 {\dot{h}_{ij}}^2 - \frac12 {h_{jk,i} }^2 + 2 \dot{h}_{ij,i} h_{0j} 
 - h_{j0,i} h_{i0,j} +h_{jk,i } h_{ik,j} -2 h_{0i} \dot{h}_{kk,i} -h_{ij,i} h_{kk,j} \nn 
&- \frac12 {\dot{h}_{ii} }^2 +\frac12 {h_{ii,j} }^2
+{h_{0j,i} }^2 - \frac{m^2}{2} [-2 {h_{0i}}^2 +{h_{ij} }^2 -{h_{ii} }^2 ]  \, ,\\
\label{OAN2C}
\mathcal{G}(h_{ij}) =& -h_{ij, ij} + h_{kk ,jj} -m^2 h_{ii}  \, .
\end{align}
Then the conjugate momenta are given by,
\begin{align}
\label{OAN3}
\pi_{00} =& \frac{\partial \mathcal{L}_0 }{ \partial \dot{h}_{00} } 
= 2 \alpha A_0  \, , \quad 
\pi_{0i} = \frac{\partial \mathcal{L}_0 }{ \partial \dot{h}_{0i} } 
= - 2 \alpha A_i \, , \quad
\pi_{ij} = \frac{\partial \mathcal{L}_0 }{ \partial \dot{h}_{ij} } = \dot{h}_{ij} 
 - \dot{h}_{kk} \delta_{ij} - \partial_{i} h_{j 0} - \partial_{j} h_{i 0} 
+2 \partial_k h_{0k} \delta_{ij}  \, , \nn 
\pi_0 =& \frac{\partial \mathcal{L}_0 }{ \partial \dot{A}_{0} } 
=  - 2 \alpha h_{00} \, , 
\quad \pi_i = \frac{\partial \mathcal{L}_0 }{ \partial \dot{A}_{i} } 
= 2 \alpha h_{0i} \, .
\end{align}
Then we find 
\begin{align}
\label{OAN4}
\dot{h}_{ij} = \pi_{ij} - \frac{1}{2} \pi_{kk} \delta_{ij} + \partial_{i} h_{j 0}
 + \partial_{j} h_{i 0}\, .
\end{align}
Equations in (\ref{OAN3}) give the following primary constraints 
\begin{align}
\label{OAN5}
\phi^1 = \pi_{00} - 2 \alpha A_0 \, , \quad 
\phi^2_i = \pi_{0i} + 2 \alpha A_i \, ,\quad 
\phi^3 = \pi_0 + 2 \alpha h_{00} \, , \quad 
\phi^4_i = \pi_i - 2 \alpha h_{i0} \, .
\end{align}
The non-vanishing components of the Poisson brackets between the 
constraints are given by
\begin{align}
\label{OAN6}
\{ \phi^1 (\vec{x}) , \phi^3 (\vec{y}) \} = - 4 \alpha \delta (\vec{x}-\vec{y}) \, ,  \quad 
\{ \phi^2_i (\vec{x} ) , \phi^4_j (\vec{y}) \} = 4\alpha \delta_{ij} \delta(\vec{x} - \vec{y}) 
\, .
\end{align}
This tells $\det\{ \phi,\phi \}\neq 0$ and we can determine the Lagrange multipliers 
and we find there are no secondary constraint. 
Then we have totally 8 constraints in the phase space. 
Because the symmetric tensor has 10 degrees of freedom and the vector has 4, 
we have originally 28 degrees of freedom in the phase space. 
By subtracting 8 degrees of freedom from the constraints, there remain 20 
degrees of freedom in the phase space, that is, 10 degrees of freedom in the 
coordinate space. 

We now investigate what could be the ten physical degrees of freedom. 
We should also note that the free part $\mathcal{L}_0$ of the Lagrangian (\ref{hhh10}) 
can be diagonalized as in (\ref{RR2}) by the redefinition, 
\be
\label{RRR1}
h_{\mu\nu} (x) = l_{\mu\nu} (x) + 4 \alpha \int d^4 y 
{\hat D}^m_{\mu\nu,\rho\sigma} (x-y) \partial^\rho A^\sigma (y) \, ,
\ee
as follows,
\begin{align}
\label{RRR2}
\mathcal{L}_0 = & - \frac{1}{2} \left( l \Box l - l^{\mu\nu} \Box l_{\mu\nu} 
 - l \partial^\mu \partial^\nu l_{\mu\nu} - l_{\mu\nu} \partial^\mu \partial^\nu l 
+ 2 l_\nu^{\ \rho} \partial^\mu \partial^\nu l_{\mu\rho} \right) \nn
& + \frac{m^2}{2} \left( l^2 - l_{\mu\nu} l^{\mu\nu} \right) 
 - \frac{4\alpha^2}{m^2} \left\{ A^\mu \Box A_\mu + \left( \partial^\mu A_\mu \right)^2 
\right\} + \frac{16\alpha^2}{3m^2} \partial^\mu A_\mu 
\left( 1 - \frac{\Box}{m^2} \right) \partial^\nu A_\nu \, .
\end{align}
In (\ref{RRR1}), ${\hat D}^m_{\mu\nu,\rho\sigma} (x-y)$ is the propagator expressed by 
coordinates $x$ and $y$ and defined by 
\begin{align}
\label{PPP1}
& \left( \eta^{\mu\nu} \eta^{\rho\sigma} \Box 
 - \frac{1}{2} \left( \eta^{\mu\rho} \eta^{\nu\sigma} + \eta^{\mu\sigma} \eta^{\nu\rho} 
\right) \Box 
 - \eta^{\mu\nu} \partial^\rho \partial^\sigma 
 - \eta^{\rho\sigma} \partial^\mu \partial^\nu l 
+ \frac{1}{2}\left( \eta^{\mu\rho} \partial^\nu \partial^\sigma 
+ \eta^{\mu\sigma} \partial^\nu \partial^\rho
+ \eta^{\nu\rho} \partial^\mu \partial^\sigma 
+ \eta^{\nu\sigma} \partial^\mu \partial^\rho \right) \right. \nn
& \left. + m^2 \left( \eta^{\mu\nu} \eta^{\rho\sigma}  
 - \frac{1}{2} \left( \eta^{\mu\rho} \eta^{\nu\sigma} + \eta^{\mu\sigma} \eta^{\nu\rho} 
\right) \right) \right) {\hat D}^m_{\rho\sigma,\alpha\beta} (x-y)
= \frac{1}{2} \left( \delta^\mu_{\ \alpha} \delta^\nu_{\ \beta} 
+ \delta^\mu_{\ \beta} \delta^\nu_{\ \alpha} \right) 
\delta ( x - y ) \, ,
\end{align}
which is given by the Fourier transformation of 
$D^m_{\alpha\beta,\rho\sigma}$ in (\ref{hh1}), 
\be
\label{RRR3}
{\hat D}^m_{\mu\nu,\rho\sigma} (x-y) = \int \frac{d^4 p}{\left( 2 \pi \right)^2} 
D^m_{\alpha\beta,\rho\sigma} \e^{ip(x-y)}\, .
\ee
Then we find
\be
\label{Q1}
h_{\mu\nu} (x) = l_{\mu\nu} (x) - \frac{2 \alpha}{m^2} 
\left( \partial_\nu A_\mu (x) + \partial_\mu A_\nu (x) 
 - \frac{2}{3} \eta_{\mu\nu} \partial_\rho A^\rho (x)
 - \frac{4}{3m^2} \partial_\mu \partial_\nu \partial_\rho A^\rho (x) \right)\, ,
\ee
which gives
\be
\label{Q2}
h = h^\mu_{\ \mu} 
= l + \frac{4\alpha}{3m^4}\left( 2 \Box + m^2 \right) \partial_\rho A^\rho \, .
\ee
The Lagrangian (\ref{RRR2}) is the sum of the Lagrangian of the Fierz-Pauli massive 
gravity and the vector field $A_\mu$ except the last term. 
The last term might be regarded to be a gauge fixing term. 
%%%%%%%%%%%%%%%%%%%
The higher derivative part can be further rewritten by using a new vector field $V_\mu$ 
as follows
\be
\label{RRR4}
\partial^\mu A_\mu \left( 1 - \frac{\Box}{m^2} \right) \partial^\nu A_\nu 
\sim \partial^\mu A_\mu \partial^\nu A_\nu + \partial^\mu V_\mu \partial^\nu A_\nu 
 - \frac{m^2}{4} V^\mu V_\mu\, .
\ee
In fact, by the variation of $V_\mu$ gives 
$V_\mu = - \frac{2}{m^2} \partial_\mu \partial^\nu A_\nu$. By substituting the    
expression of $V_\mu$, we obtain the original expression. 
We now define a propagator $\Delta_{\mu\nu}$ by 
\be
\label{RRR5}
\left( \eta^{\mu\nu} \Box + \frac{1}{3} \partial^\mu \partial^\nu \right)
\Delta_{\nu\rho} (x-y) = \delta^\mu_{\ \rho} \delta (x-y)\, .
\ee
Then redefining $A_\mu$ by 
\be
\label{RRR6}
A_\mu = B_\mu - \frac{2}{3} 
\int d^4 y \Delta_{\nu\rho} (x-y) \partial^\rho \partial^\sigma V_\sigma (y)\, ,
\ee
the Lagrangian density (\ref{RRR2}) can be rewritten as
\begin{align}
\label{RRR7}
\mathcal{L}_0 = & - \frac{1}{2} \left( l \Box l - l^{\mu\nu} \Box l_{\mu\nu} 
 - l \partial^\mu \partial^\nu l_{\mu\nu} - l_{\mu\nu} \partial^\mu \partial^\nu l 
+ 2 l_\nu^{\ \rho} \partial^\mu \partial^\nu l_{\mu\rho} \right) \nn
& + \frac{m^2}{2} \left( l^2 - l_{\mu\nu} l^{\mu\nu} \right) 
 -  \frac{4\alpha^2}{m^2} \left\{ B^\mu \Box B_\mu 
+ \left( \partial^\mu B_\mu \right)^2 \right\} 
+ \frac{16\alpha^2}{3m^2} \left( \partial^\mu B_\mu \right)^2 
 - \frac{4\alpha^2}{3m^2} \left( \partial^\mu V_\mu \right)^2  
 - \frac{4\alpha^2}{3} V^\mu V_\mu \, .
\end{align}
The Lagrangian density is the sum of the Lagrangian of the Fierz-Pauli massive 
spin two field $l_{\mu\nu}$ and the vector field $B_\mu$ with gauge fixing term and the 
action of an exotic vector field $V_\mu$. 
Because $V_i$'s are not dynamical but $V_0$ is dynamical because there is no term 
including the derivative of $V_i$'s with respect to time. 
Therefore $V_\mu$ contains only one degrees of freedom. 
Because $A_\mu$ has four degrees of freedom after the gauge fixing, we 
have totally ten degrees of freedom including the massive graviton $l_{\mu\nu}$, 
which is consistent with the previous Hamiltonian analysis. 

In the Lagrangian density (\ref{RRR7}), the sign in front of the kinetic term of the 
vector field is not canonical and therefore the vector field is ghost. 
Although the model contains ghost fields, the model could be renormalizable and 
therefore the model proposed in this paper might be regarded as a kind of toy model. 
If we could extend the model to have a local symmetry, some physical state condition 
may select physical states where no ghost state appears. 

The bigravity model can be regarded as a model where massive spin two field couples with 
gravity. 
Then we may consider the model where $h_{\mu\nu}$, whose Lagrangian is given by 
(\ref{hh10}) couples with gravity
\begin{align}
\label{hhhh1}
S =& \int d^4 x \sqrt{-g} \left\{ 
 - \frac{1}{2} g^{\mu_{1} \nu_{1} \mu_{2} \nu_{2} \mu_{3} \nu_{3}} 
\nabla_{\mu_{1}} \nabla_{\nu_{1}} h_{\mu_{2} \nu_{2}} h_{\mu_{3} \nu_{3}}
+ \frac{1}{2} m^2 g^{\mu_{1} \nu_{1} \mu_{2} \nu_{2}} h_{\mu_{1} \nu_{1}} h_{\mu_{2} \nu_{2}} 
\right. \nn
& \left. - \frac{\mu}{3!} g^{\mu_{1} \nu_{1} \mu_{2} \nu_{2} \mu_{3} \nu_{3}} 
h_{\mu_{1}\nu_{1}} h_{\mu_{2} \nu_{2}} h_{\mu_{3} \nu_{3}}
 - \frac{\lambda}{4!} g^{\mu_{1} \nu_{1} \mu_{2} \nu_{2} \mu_{3} \nu_{3} \mu_{4} \nu_{4}} 
h_{\mu_{1} \nu_{1}} h_{\mu_{2} \nu_{2}} h_{\mu_{3} \nu_{3}} h_{\mu_{4} \nu_{4}}
%+ 4\alpha A^\mu \nabla^\nu h_{\mu\nu} 
\right\} \, ,
\end{align}
which can be regarded as a new bigravity model because there appear two symmetric 
tensor fields $g_{\mu\nu}$ and $h_{\mu\nu}$. 
We should note that $h_{\mu\nu}$ is not the perturbation in $g_{\mu\nu}$ but 
$h_{\mu\nu}$ is a field independent of $g_{\mu\nu}$. 
%%%%%%%%%%
Because the gravity is not renormalizable, we forget about the renormalizability and 
drops the last term in (\ref{hhh10}), where the vector field $A_\mu$ couples with 
$h_{\mu\nu}$. 
%%%%%%%%%%

%%%%%%%%%%%%%%%%%%
\section{Summary}

In summary, we considered the non-linear derivative interactions which are not 
included in \cite{deRham:2013tfa} but unfortunately we have shown that such derivative 
interactions could generate ghost. 
We also investigated the possibility of other classes of the no-ghost interactions by 
only requiring the Lorentz invariance. 

Motivated with the above analyses, we proposed a power counting renormalizable model 
describing the massive spin two field, which could not be really 
renormalizable because the projection operators included in the propagator 
generate non-renormalizble divergences. 
We solved this problem by adding a new term where a vector field $A_\mu$ couples 
with the massive spin two field $h_{\mu\nu}$. 
By investigating the spectrum of this model, it was shown that there could appear ghost 
and therefore the model cannot be realistic one but we can regard this model as 
a kind of toy model, which may be a candidate of the renormalizable model. 

Because the gravity is not renormalizable, we may consider the coupling of the power 
counting renormalizable model, which could not be really renormalizable, with gravity. 
The model can be regarded as a new kind of bimetric 
gravity or bigravity. 
In the Appendix, we have shown that the field of the massive spin two 
field plays the role of the cosmological constant. 
It is easy to see that the vacuum solution like the Schwarzschild solution or Kerr solution 
in the Einstein gravity becomes a solution of the new bigravity model. 

\ 

\noindent
{\bf Aknowledgments} We are grateful to S.~D.~Odintsov for useful discussions. 
We are also indebted to T.~Maskawa for the suggestions about the massive vector field. 
The work is supported by the JSPS Grant-in-Aid for Scientific 
Research (S) \# 22224003 and (C) \# 23540296 (S.N.) 
and that for Young Scientists (B) \# 25800136 (K.B.).  

\appendix

\section{Cosmology by new bigravity}

We may consider the cosmology given by the action (\ref{hhhh1}) with the Einstein-Hilbert 
action:
\be
\label{EH}
S_\mathrm{EH} = \frac{1}{2\kappa^2} \int d^4 x \sqrt{-g} R\, .
\ee
We assume the solution of equations given by the actions (\ref{hhhh1}) and (\ref{EH}) is 
given by 
\be
\label{hhhh2}
h_{\mu\nu} = C g_{\mu\nu}\, .
\ee
Here $C$ is a constant. We can directly check that Eq.~(\ref{hhhh2}) satisfies the 
field equation given by the variation of $h_{\mu\nu}$ and also the Einstein 
equation by properly choosing $C$. 
By substituting (\ref{hhhh2}) into the action (\ref{hhhh1}), we find
\be
\label{hhhh3}
S = - \int d^4 x \sqrt{-g} V(C)\, ,\quad 
V(C) \equiv - 6m^2 C + 4\mu C^3 + \lambda C^4\, .
\ee
We should note $\nabla_\rho g_{\mu\nu}=0$. 
The constant $C$ can be determined by the equation $V'(C)=0$. 
We now parametrize $m^2$ and $\mu$ by
\be
\label{hhhh4}
m^2 = - \frac{\lambda}{3} C_1 C_2\, ,\quad 
\mu = - \frac{\lambda}{3} \left( C_1 + C_2 \right)\, .
\ee
Then the solutions of $V'(C)$ are given by 
\be
\label{hhhh5}
C=0,\, C_1,\, C_2\, ,
\ee
and we find
\be
\label{hhhh6}
V\left(C_1\right) = \frac{\lambda}{3} C_1^3\left( - C_1 + 2 C_2 \right) \, , \quad 
V\left(C_2\right) = \frac{\lambda}{3} C_2^3\left( - C_2 + 2 C_1 \right) \, . 
\ee
Then we find $V(C)$ plays the role of cosmological constant. 
Let assume $0<C_1<C_2$ and $C_2<2C_1$. 
Then $V\left(C_1\right)$ is a local maximum and $V\left(C_2\right)>0$ is a 
local minimum. 
Then $V\left(C_1\right)$ or $V\left(C_2\right)$ might generate the inflation. 

It has been shown that the causality could be broken in the previous bigravity models 
\cite{Deser:2013eua} due to the existence of the superluminal mode. 
We should note that in the model given by the actions (\ref{hhhh1}) and (\ref{EH}), 
the superluminal mode does not appear and therefore the causality could not be 
broken. 

We should also note that under the assumption (\ref{hhhh2}), we can construct 
black hole solutions as in the standard bigravity model (see, for example 
\cite{Katsuragawa:2013bma,Katsuragawa:2013lfa}).

\end{document}